\pdfoutput=1
\documentclass[12pt]{article}

\usepackage[pdftex]{graphicx}
\usepackage{graphicx}
\usepackage{amsmath}
\usepackage{amsfonts}
\usepackage{amssymb}
\usepackage{bbm}
\usepackage{appendix}

\begin{document}

\newcommand{\ba}[1]{\begin{array}{#1}} \newcommand{\ea}{\end{array}}

\numberwithin{equation}{section}


\def\Journal#1#2#3#4{{#1} {\bf #2}, #3 (#4)}

\def\NCA{\em Nuovo Cimento}
\def\NIM{\em Nucl. Instrum. Methods}
\def\NIMA{{\em Nucl. Instrum. Methods} A}
\def\NPB{{\em Nucl. Phys.} B}
\def\PLB{{\em Phys. Lett.}  B}
\def\PRL{\em Phys. Rev. Lett.}
\def\PRD{{\em Phys. Rev.} D}
\def\ZPC{{\em Z. Phys.} C}

\def\st{\scriptstyle}
\def\sst{\scriptscriptstyle}
\def\mco{\multicolumn}
\def\epp{\epsilon^{\prime}}
\def\vep{\varepsilon}
\def\ra{\rightarrow}
\def\ppg{\pi^+\pi^-\gamma}
\def\vp{{\bf p}}
\def\ko{K^0}
\def\kb{\bar{K^0}}
\def\al{\alpha}
\def\ab{\bar{\alpha}}

\def\np{Nucl. Phys. {\bf B}}\def\pl{Phys. Lett. {\bf B}}
\def\mpl{Mod. Phys. {\bf A}}\def\ijmp{Int. J. Mod. Phys. {\bf A}}
\def\cmp{Comm. Math. Phys.}\def\prd{Phys. Rev. {\bf D}}

\def\oa{\bigcirc\!\!\!\! a}
\def\ob{\bigcirc\!\!\!\! b}
\def\oc{\bigcirc\!\!\!\! c}
\def\oi{\bigcirc\!\!\!\! i}
\def\oj{\bigcirc\!\!\!\! j}
\def\ok{\bigcirc\!\!\!\! k}
\def\ve{\vec e}\def\vk{\vec k}\def\vn{\vec n}\def\vp{\vec p}
\def\vr{\vec r}\def\vs{\vec s}\def\vt{\vec t}\def\vu{\vec u}
\def\vv{\vec v}\def\vx{\vec x}\def\vy{\vec y}\def\vz{\vec z}

\def\ve{\vec e}\def\vk{\vec k}\def\vn{\vec n}\def\vp{\vec p}
\def\vr{\vec r}\def\vs{\vec s}\def\vt{\vec t}\def\vu{\vec u}
\def\vv{\vec v}\def\vx{\vec x}\def\vy{\vec y}\def\vz{\vec z}

\newcommand{\AdS}{\mathrm{AdS}}
\newcommand{\dd}{\mathrm{d}}
\newcommand{\eee}{\mathrm{e}}
\newcommand{\sgn}{\mathop{\mathrm{sgn}}}

\def\a{\alpha}
\def\b{\beta}
\def\g{\gamma}

\newcommand\lsim{\mathrel{\rlap{\lower4pt\hbox{\hskip1pt$\sim$}}
    \raise1pt\hbox{$<$}}}
\newcommand\gsim{\mathrel{\rlap{\lower4pt\hbox{\hskip1pt$\sim$}}
    \raise1pt\hbox{$>$}}}

\newcommand{\beq}{\begin{equation}}
\newcommand{\eeq}{\end{equation}}
\newcommand{\bea}{\begin{eqnarray}}
\newcommand{\eea}{\end{eqnarray}}
\newcommand{\noi}{\noindent}


\begin{flushright}
March, 2012
\end{flushright}

\bigskip

\begin{center}

{\Large\bf  Spontaneous breaking of a discrete symmetry and holography}
\vspace{1cm}

\centerline{Borut Bajc$^{a,b,}$\footnote{borut.bajc@ijs.si}, Adri\'{a}n R. Lugo$^{c,}
$\footnote{lugo@fisica.unlp.edu.ar} and Mauricio B. Sturla$^{d,}$\footnote{sturla@icmm.csic.es}}

\vspace{0.5cm}
\centerline{$^{a}$ {\it\small J.\ Stefan Institute, 1000 Ljubljana, Slovenia}}
\centerline{$^{b}$ {\it\small Department of Physics, University of Ljubljana, 1000 Ljubljana, Slovenia}}
\centerline{$^{c}$ {\it\small Departamento de F\'\i sica and IFLP-CONICET, }}
\centerline{ {\it\small Facultad de Ciencias Exactas, Universidad Nacional de La Plata,}}
\centerline{ {\it\small  C.C. 67, 1900 La Plata, Argentina}}
\centerline{$^{d}$ {\it\small Instituto de Ciencia de Materiales de Madrid, CSIC,}}
\centerline{{\it\small Cantoblanco, E-28049, Madrid, Spain.}}

\end{center}

\bigskip

\begin{abstract}
We present an exactly solvable  model of a scalar field in an AdS$_{d+1}$
like background interpolating between a $Z_2$
preserving and a $Z_2$ breaking minima of the potential. We define its holographic dual through the AdS/CFT
dictionary and argue that at zero temperature the $d-$dimensional strongly coupled system on the boundary
of AdS$_{d+1}$ exhibits a phase with a spontaneously broken discrete symmetry.
In the presence of a black hole in the bulk ($T\neq 0$) we find that, although the metastable phase
is present, the discrete symmetry gets restored.
We compute exactly the lowest order boundary correlation functions in the spontaneously broken phase at
$T=0$, finding out a pole of the propagator for zero momenta that signals the presence of a massless mode 
and argue that it should not be present at $T\neq 0$. 

\end{abstract}

\clearpage

\tableofcontents

\section{Introduction}

The goal of this work is to study the behavior of a symmetry in a strongly coupled
system in d-dimensional space ($M_d$) using the AdS/CFT correspondence \cite{malda,gubser,witten}.
We will be interested in the case of a $Z_2$ discrete symmetry which can be considered the prototype
example for larger discrete, global or local symmetries.
In doing so
we will develop some machinery and a controlled systematic approximation which allows analytic results,
and is by itself worth presenting.

While originally the study of the AdS/CFT conjecture focused mainly on supersymmetric theories
with known (or guessed) dual supersymmetric quantum field theories (QFT) \cite{D'Hoker:2002aw},
since the seminal work in reference \cite{Hartnoll:2008vx} much effort has been put in the last years on
applications of the duality to the study of condensed matter systems \cite{Hartnoll:2009sz}.
In this kind of applications, the correspondence is used just as a dictionary to get properties
(phase diagrams, correlation functions, transport coefficients, etc.) of a strongly coupled QFT,
whose lagrangean formulation is in general unknown, from the knowledge of its gravitational dual.
This is the point of view we adopt in this paper.

We will study a real scalar field with usual canonical kinetic term plus a potential with a
$Z_2$ symmetry and three locally stationary points - minima
\footnote{
In principle a local maximum can be allowed as a classically stable point providing the negative
mass square satisfies the Breitenl\"ohner-Freedman bound \cite{Breitenlohner:1982jf}
}.
The $Z_2$ symmetry in the bulk as the ancestor of a $Z_2$ symmetry on the boundary theory is
motivated by the notion of gauged discrete symmetries \cite{Krauss:1988zc}, and as such resembles
the usual correspondence of a gauge symmetry in the bulk to a global symmetry on the boundary.
In principle such a bulk theory could be seen for example as a low energy limit of a
$U(1)$ gauge theory spontaneously broken by a doubly charged Higgs vacuum expectation value (vev).

We give emphasis on exact results, reason that leads us to present a soluble, although non trivial,
piece-wise quadratic potential in a fixed (i.e. with no back-reaction) AdS background.
We will show that a spontaneous symmetry breaking (SSB) solution exists  at non zero temperature
for any value of the parameters of the potential, but with a higher free energy density than
the symmetric phase solution, representing thus a kind of metastable phase.
Instead, at $T=0$ the solution exists with equal energy, representing a genuine phase where
SSB is realized. In this case a non-trivial constraint among the potential parameters is needed
for the solution to exist. This same relation has at the same time a far reaching physical consequence:
it leads to the existence of a massless state in the boundary QFT.

The paper is organized as follows.
In Section $2$ we present the general set-up; in Section $3$ we specialize to a simple although non
trivial model which can be exactly solvable, showing the existence of the SSB solution; in Section $4$ we analyze two and three point correlation functions. Finally we include two appendices; in the first one we derive a formula for the free energy of the system, while in the second one we present a formalism to compute correlation functions in the AdS/CFT framework.

\section{The set-up}

We consider a real scalar field $\phi$ in $d+1$ dimensions with bulk euclidean action

\beq
\label{action}
S^{(bulk)}[\phi] =  \int d^{d+1}x\,\sqrt{|\det{g_{ab}}|}\;
\left(\frac{1}{2}\partial_a\phi\; g^{ab}\;\partial_b\phi+U(\phi)\right)
\eeq
in a non-dynamical $AdS_{d+1}$ black hole background
\beq\label{ads}
ds^2 = \frac{L^2dz^2}{z^2\left(1-(z/z_h)^d\right)}
+ \frac{1}{z^2}\;\left(d\vec x^2 + \left(1-(z/z_h)^d\right)\;d\tau^2\right)
\eeq
where $x=(\vec x,\tau)$ are the QFT coordinates with $\tau$ the euclidean time with
periodicity, $\tau\rightarrow \tau + 1/T$.
While the boundary is at $z=0$, the horizon is fixed at $z=z_h>0$, so that
the Hawking temperature is
\beq
T=\frac{d}{4\,\pi\, L\,z_h}
\eeq
For $\phi=\phi(z)$ the equation of motion derived from (\ref{action}) results
\beq
\label{eom}
\left(1-(z/z_h)^d\right)z^2\;\phi''(z) -
\left(d-1+(z/z_h)^d\right)z\; \phi'(z) =L^2\;U'(\phi)
\eeq
where the prime indicates the derivative w.r.t. the bulk coordinate $z$.
Due to the $Z_2$ symmetry, it is enough to consider the potential for
positive values of the field $\phi$ only. The form of the potential we will be
interested in is shown in fig. 1. Denoting the true minimum with $\phi_m$ (the
false vacuum is at $\phi=0$) and the local maximum with $\phi_M$,
we can redefine the field variable as the dimensionless
\beq
t(z)\equiv\frac{\phi(z)}{\phi_m}\qquad;\qquad t_M = \frac{\phi_M}{\phi_m}\quad,\quad t_m =1
\eeq
and the potential as
\beq
V(t)\equiv \frac{L^2}{\phi_m^2}U(\phi_m t)
\eeq
The equation of motion to solve is then
\beq\label{eqfortgeneral}
\left(1-(z/z_h)^d\right)z^2\;t''(z) -
\left(d-1+(z/z_h)^d\right)z\; t'(z) =  V'(t)
\eeq
In the following we will consider the two cases: $T=0$ ($z_h=\infty$) and
$T\ne 0$ ($z_h=1$).

\subsection{\label{adsqft}The AdS/QFT interpretation}

In general, the equation of motion in an asymptotic AdS$_{d+1}$ background (in particular, (\ref{ads}))
implies that at small $z$ (close to the boundary) the scalar field must behave as
\bea\label{bb}
\phi(x,z) &\xrightarrow{z\rightarrow 0}& \phi_v + \phi_{(0)}(x)\;z^{\Delta_-}
(1+ \dots) +\phi_{(1)}(x)\;z^{\Delta_+}(1 + \dots)\cr
\Delta_\pm &=& \frac{d}{2}\pm\sqrt{\frac{d^2}{4} + L^2\,U''(\phi_v)}
\eea
where the dots stand for corrections with positive powers  of $z$. We will avoid terms proportional
to $\log{z}$ by considering non-integer values for $\Delta_\pm$. $\phi_v$ is a stationary
point of the potential, i.e.  $U'(\phi_v)=0$. For our case
the interesting choice is $\phi_v=0$, which we will keep from now on.
If we choose a fixed but arbitrary $\phi_{(0)}(x)$ at the boundary, the solution to the equation of motion
then determine $\phi_{(1)}(x)$ as a functional of $\phi_{(0)}(x)$.

The AdS/QFT conjecture states that given any bulk field $\phi(x,z)$, its fixed boundary value
field $\phi_{(0)}(x)$ acts as a source for the correlation functions of a dual operator $\hat O (x)$ of the QFT,
\beq\label{adscft}
Z_{gravity}^{(bulk)}[\phi_{(0)}] = Z_{QFT}^{(boundary)}[\phi_{(0)}] \equiv
< \exp{\left(\int_{M_d}\,\phi_{(0)}\,\hat O\right)} >
\eeq
We will assume that in some limit (in string theory this can be a large $N$ limit) the bulk partition function can
be computed from its classical on-shell action:
\beq\label{genfunct}
Z_{gravity}^{(bulk)}[\phi_{(0)}] = \exp{\left(-\left.S[\phi]\right|_{\it on-shell}\right)}
\eeq
Then $-S[\phi]|_{\it on-shell}$ can be considered as the generating functional of
connected correlation functions of the operator $\hat O(x)$. It is made out of the bulk
action (\ref{action}) properly renormalized with boundary terms that make it
finite and give sense to the variational problem, see for example \cite{Skenderis:2002wp}.

We can further simplify the information in (\ref{genfunct}) by the well known fact in field theory
that the solution of the classical equation of motion in the presence of a source is the functional derivative
over the source of the generating functional for tree level connected diagrams.
Since the relevant part of the functional dependence of the variation of the action on $\phi_{(0)}(x)$ comes
through integration by parts and boundary terms, only the limiting part at $z\to 0$ of the solution
to the equation of motion enters the game \cite{Hartnoll:2008vx}:
\beq\label{vev}
< \hat O(x)>_{\phi_{(0)}}= -\frac{\delta S[\phi]|_{\it on-shell}}{\delta\phi_{(0)}(x)} =
\frac{\Delta_+ -\Delta_-}{L}\;\phi_{(1)}(x)
\eeq
where the subindex of the vev reminds us that the above is meant in the presence of
an arbitrary source $\phi_{(0)}$. Higher point correlation functions are then evaluated by further functional derivatives.

\subsection{\label{freeenergy}The free energy}

Each solution with fixed boundary conditions represents a phase of the boundary theory, and the more favored
one will have less free energy density $f$.
According to (\ref{adscft}) it is given by,
\beq\label{fe1}
f \equiv \frac{T}{V_{d-1}}\; S[\phi]
\eeq
In our model, the vacuum solution $\phi = \phi_v=0$ is dual to the symmetric phase of the QFT and has
$f = 0$. Any non trivial solution will signal the existence of a phase where the spontaneous breakdown of
the $Z_2$ symmetry is present: a negative sign of $f$ would mean that the broken phase is preferable,
a vanishing $f$ that it can coexist with the unbroken phase, and a positive one that it is metastable.

We can say something about the on-shell action for a general continuous potential $V$
even without solving the problem explicitly. Let us write,
\bea
f &=& \frac{\phi_m{}^2}{L}\; I [t_b]\cr
I[t]&\equiv&\int_0^{z_h} \frac{dz}{z^{d+1}}\left(\frac{z^2}{2}(1-(z/z_h)^d)\;t'^2(z)+V(t(z))\right)
\eea
Let us assume that there exists a well-behaved solution $t_b$, continuous and with continuous derivative,
that goes to zero for $z\rightarrow 0$.
By replacing one $z\,t'(z)$ above from the equation of motion
and integrating by parts we get (see Appendix A),
\bea\label{actionfinal}
I[t_b]&=&\left[\frac{(1-(z/z_h)^d)}{d\,z^d}\left(\frac{z^2}{2}(1-(z/z_h)^d)\;t_b'^2-V(t_b)\right)\right]_{z=0}^{z=z_h}\nonumber\\
&+&\frac{1}{2}\int_0^{z_h}dz\; z^{1-d}\;(z/z_h)^d\;(1-(z/z_h)^d)\;t_b'^2
\eea
In the region $z\to 0$ the requirement for a
(vanishing) local minimum means that $V\propto t^2$, and since $t\propto z^{\Delta_+}$,
the contribution of the lower limit of the first line to the action goes like $z^{2\,\Delta_+ -d}$ and so is zero.
In the other limit, $z\to z_h$, the potential again goes to a finite value, while $|t'|<\infty$.
So the boundary and horizon limits do not contribute for any $T$.

This means that {\it at $T=0$} ($z_h=\infty$) {\it the total action on the solution is vanishing
for any potential $V$} of the form we are considering:
\beq
f (T=0)=0
\eeq
For a non-zero temperature ($z_h=1$) on the other side the action is strictly positive
for any non-trivial solution:
\beq
f(T\ne 0) = \frac{\phi_m{}^2}{2\,L}\;\int_0^{1}dz\; z\;(1-z^d)\;t_b'(z)^2\;>0
\eeq

We can thus already conclude that if a SSB solution exists, it is metastable at $T\ne 0$ but energetically
equivalent to the symmetric one at $T=0$.

\section{A piece-wise quadratic potential: the exact solution}

We present in this Section a solvable, non trivial model which allows exact solutions
\footnote{
Models with simplified potentials as the one considered here were already studied in other contexts,
see for example \cite{refpotsimp}.}.

The interesting region for $t$ is between the local minimum at $0$ and
the global minimum at $1$.
We will divide this region into a number of sections, and in each of them the potential can
be locally approximated by a quadratic form:
\beq
\label{vt}
V(t)=\frac{A}{2}\;t^2+B\;t+C
\eeq
The minimum number of such sections is three: (1) $0<t<t_1$, (2) $t_1<t<t_2$, (3) $t_2<t<1$.
The coefficients in (\ref{vt}) are parameterized in each region as
\bea
\label{ABC}
A &=& \left\{\begin{array}{r}A_1>0\\A_2<0\\A_3>0 \end{array}\right.
\qquad;\qquad B = \left\{ \begin{array}{l} 0\\-A_2\,t_M\\-A_3\end{array}\right.
\qquad;\qquad\cr
C &=& \left\{ \begin{array}{l}0\\(A_1-A_2)\;t_1{}^2/2+A_2\;t_M\;t_1\cr
(A_2-A_3)\;t_2{}^2/2+A_3\;t_2-A_2\;t_M\;t_2+(A_1-A_2)\;t_1{}^2/2+A_2\;t_M\;t_1 \end{array}\right.
\eea
respectively.
The strange choice of $C$'s is required by the continuity of the potential.
Furthermore, we will require the continuity of the first derivatives of the
potential which yields to
\beq\label{contderpot}
t_1 = \frac{-A_2\;t_M}{A_1 - A_2}\qquad;\qquad
t_2 = \frac{A_3-A_2\,t_M}{A_3-A_2}
\eeq
relations that automatically satisfy $0<t_1<t_M<t_2<1$ for any $0<t_M<1$.
In this way we remain with four relevant parameters, the $A_i$'s and $t_M$.

For further use we also introduce
\beq\label{mus}
\mu_i{}^2 \equiv \frac{1}{4}+ \frac{A_i}{d^2}\qquad;\qquad \Delta_i^\pm\equiv d\left(\frac{1}{2}\pm\mu_i
\right)
\eeq
We will consider the case of real $\mu_{1,3}>\frac{1}{2}$ ($A_{1,3}>0$) and pure imaginary
$\mu_2\equiv i\,\bar\mu_2$ ($A_2<-\frac{d^2}{4}$) with $\bar\mu_2>0$.

\begin{figure}[htb]
\begin{center}
\includegraphics[width=10.cm]{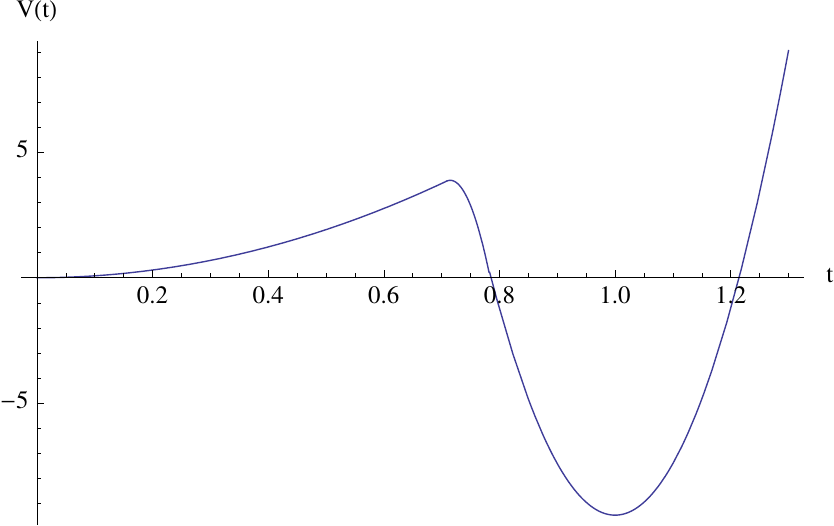}
\caption{\label{fig1} The potential for
$\mu_1=1.1$, $\bar\mu_2=10.1$, $\mu_3=5.1$ and $t_M=0.715$ ($k=0$).}
\end{center}
\end{figure}

An example of such a potential is shown in fig. \ref{fig1}.

\subsection{The solution for $T = 0$}

The scalar equation of motion (\ref{eqfortgeneral}) for the background solution $t_b(z)$ simplifies to
\beq\label{em4}
z^2\,t_b''(z)-(d-1)z\,t_b'(z)- A \, t_b(z)= B
\eeq
The solution to (\ref{em4}) we are looking for can be written as
\beq
\label{tb}
t_b(z)=\left\{
\begin{array}{lcr}
t_1\; (z/z_1)^{\Delta_1^+} & , &  0<z<z_1\\
t_M+D_+\,(z/z_2)^{\Delta_2^+}+D_-\,(z/z_2)^{\Delta_2^-}& , & z_1<z<z_2\\
1-(1-t_2)\,(z/z_2)^{\Delta_3^-} & , &  z_2<z<\infty
\end{array}
\right.
\eeq
From the continuity of the solution and its derivative at $z_{1,2}$ we get four equations.
In principle they should determine the four unknowns $D_{+,-}$ and $z_{1,2}$, but
due to dilatation invariance $z\to \lambda z$ of (\ref{em4}) only the ratio $z_2/z_1>1$
appears. It is straightforward to show  that in spite of this, a solution is still possible for a special, quantized
choice
\beq
\label{z12cont}
d\,\bar\mu_2\,\ln{(z_2/z_1)}=(2\,k+1)\;\pi-\alpha_1-\alpha_3\qquad;\qquad k=0, 1, 2,\dots
\eeq
where we used (\ref{contderpot}) and defined the phases
\beq\label{alfai}
0<\alpha_i\equiv\arctan\frac{\bar\mu_2}{\mu_i}<\frac{\pi}{2}\quad,\quad i=1,3
\eeq
This then leads to a quantized relation among the potential parameters $t_M$, $\mu_{1,3}$ and $\bar\mu_2$:
\beq
\label{tmk}
t_M^{(k)}=\left(1+\frac{\mu_1+1/2}{\mu_3-1/2}\;
\left(\frac{\mu_3^2+\bar\mu_2^2}{\mu_1^2+\bar\mu_2^2}\right)^{1/2}
\exp{\left(\frac{(2k+1)\pi-\alpha_1-\alpha_3}{2\,\bar\mu_2}\right)}\right)^{-1}
\eeq
As we anticipated, solutions to our system of equations exist only
for special discrete values of the model parameter $t_M$.
Not all $k$ are however allowed for the solution (\ref{tb}).
In fact, what can happen is that the solution as a function of $z$ instead of increasing starts decreasing (or oscillating) at some point.
This may lead to a change of the region of the piecewise potential, so that
in such a case more intervals in $z$ should be included for the solution.

As an example, we can take $\mu_1=1.1$, $\bar\mu_2=10.1$, $\mu_3=5.1$ and $k=0$.
The output parameters are given by $z_2/z_1=1.014$, $t_1=0.708$, $t_M=0.715$ and $t_2=0.772$.
The  solution with $z_1=1$ is shown in fig. \ref{fig2}. The ansatz (\ref{tb}) is consistent also for $k=1,2$, but
not for $k>2$. In that case one should add one or more $z$-intervals.

Finally, from $(\ref{bb})$, $(\ref{vev})$ and (\ref{tb}) we read the vev of the order parameter signalling the spontaneous breakdown of the discrete symmetry,
\beq\label{vev1}
<\hat O> \;=\; \frac{2\,d\,\mu_1\,\phi_m}{L\,z_1^{\Delta_1^+}}\,t_1
\eeq
More about correlation functions will be discussed in Section \ref{corfun}.

\subsection{The solution for $T\neq 0$}

The equation of motion (\ref{eqfortgeneral}) for $z_h=1$ takes the form,
\beq
(1-z^d)\;z^2\;t_b''(z)-(d-1+z^d)\;z\;t_b'(z)- A \; t_b(z)= B
\eeq
Two independent solutions of the homogeneous part exist in any region,
\beq
\label{tpm}
t_\mu^\pm(z)=z^{d(1/2\pm\mu)}\;
F\left(\frac{1}{2}\pm\mu,\frac{1}{2}\pm\mu;1\pm 2\mu;z^d\right)
\eeq
where $\mu$ is as in (\ref{mus})
and $F\left(a,b;c;z\right)= {}_2F_1\left(a,b;c;z\right)$ is the standard hypergeometric function.
For $z\in[0,1]$, both solutions (\ref{tpm}) are real for $\mu\in\mathbb{R}$, and conjugate-related for $\mu\in i\,\mathbb{R}$.

For $z\to 0$  they behave as
\bea
t_\mu^\pm(z)\to\;z^{\Delta_\pm}\;\left(1+{\cal O}(z^{d})\right)
\eea
So in the small $z$ region (near the boundary) only $t_{\mu_1}^+(z)$ is admissible, i.e. the one that drops as $\Delta_1^+$.

The behavior for $z\to 1^-$ is on the other side like,
\beq
t_{\mu}^\pm(z)\to-\frac{\Gamma(1\pm 2\,\mu)}{\Gamma(\frac{1}{2}\pm\mu)^2}\;
\ln{(1-z)}+{\cal O}(1)
\eeq
Neither of them is admissible close to the horizon at $z=1$ but it is so the obvious
linear combination that can be chosen,
\beq
Q_{\mu_3}(z) \equiv \frac{\Gamma(1-2\mu_3)}{\Gamma(\frac{1}{2}-\mu_3)^2}\;t_{\mu_3}^+(z)-
 \frac{\Gamma(1+2\mu_3)}{\Gamma(\frac{1}{2}+\mu_3)^2}\;t_{\mu_3}^-(z)
\eeq
To closely follow the $T=0$ case, we write the solution in the form,
\beq\label{t1}
t_b(z) = \left\{ \begin{array}{lcl} t_1\; \frac{t_{\mu_1}^+(z)}{t_{\mu_1}^+(z_1)} &\quad,\quad& 0<z<z_1 \\
t_M + D_+\;t_{\mu_2}^+(z) + D_-\; t_{\mu_2}^-(z)&\quad,\quad&z_1<z<z_2\\
1 - (1- t_2)\;\frac{Q_{\mu_3}(z)}{Q_{\mu_3}(z_2)}&\quad,\quad& z_2< z< 1\end{array}\right.
\eeq
There is no dilatation symmetry here, so the number of unknowns $(D_+, D_-, z_1, z_2)$
matches the number of equations from continuity of the solution and its derivative at
$z=z_1$ and $z=z_2$. We expect to find the solution for continuous choices of the model
parameters $A_{1,2,3}$, $t_M$. Notice that at $T=0$ the input were $A_{1,2,3}$ and a flat
direction $z_1$.

\begin{figure}[htb]
\begin{center}
\includegraphics[width=6.5cm]{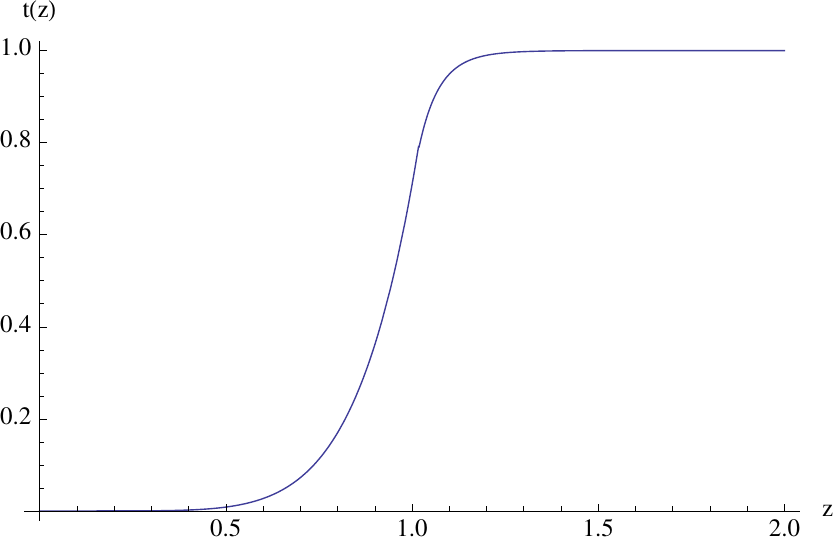}
\includegraphics[width=6.5cm]{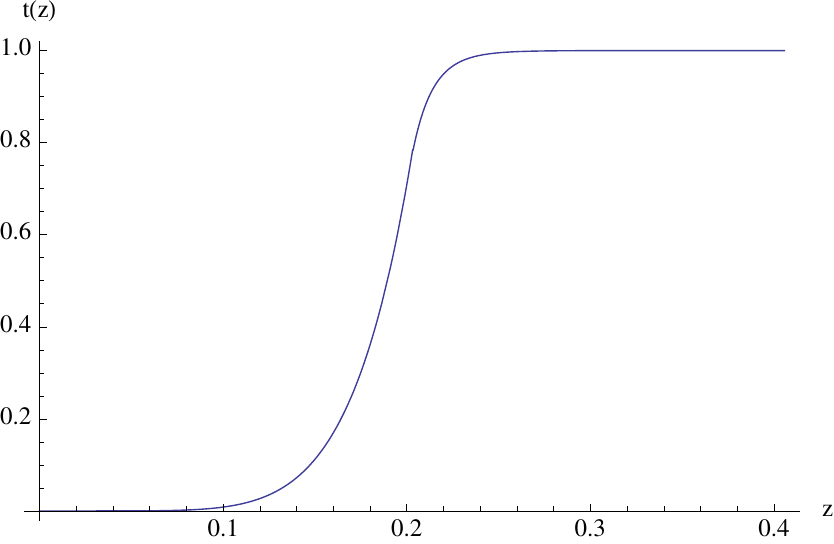}
\caption{\label{fig2} The solution for $T=0$ (left) and $T\ne 0$ (right) for
$\mu_1=1.1$, $\bar\mu_2=10.1$, $\mu_3=5.1$, $t_M=0.715$ ($k=0$). The $T=0$ (massless) wall has
been chosen to start at $z_1=1$, but can be freely translated to the left or right.}
\end{center}
\end{figure}

As an example we can consider the same input parameters that we used in the $T=0$ case:
$\mu_1=1.1$, $\bar\mu_2=10.1$, $\mu_3=5.1$. Instead of $z_1$, which is now an output rather than
an input, we choose the same $t_M$ as before, i.e. $t_M=0.715$ (at $T=0$ this was an output).
The corresponding solution gives $z_1=0.2$, $z_2=0.203$, $t_1=0.708$, $t_2=0.772$, resulting into a
positive action integral $I[t_b]=1.184$. The solution plot is given in fig. \ref{fig2}.

For $T\ne 0$ the allowed region is only
$z\in [0, 1]$, while for $T=0$ it is $z\in [0,\infty]$. The solutions for the domain wall
close to the boundary $z=0$ are very similar in the two cases, with the differences
hardly noticeable. On the contrary, when the domain wall goes towards larger values
(close to $z=1$ or bigger) the differences become more and more evident.
It is worth stressing that although at finite temperature the position of the domain wall
is fixed by the initial choice of the potential parameters (i.e. both $z_1$ and $z_2$ are
determined by the continuity of the solution), this is no longer true for
vanishing temperature, where only the ratio $z_2/z_1$ can be evaluated and thus
the domain wall position is undetermined. In any case a fundamental difference
between the two cases is, as we already mentioned, the value of the action (free energy):
vanishing for $T=0$ and positive for $T>0$, in accord with (\ref{actionfinal}).

Finally, from $(\ref{bb})$, $(\ref{vev})$ and (\ref{t1}) we read the vev of the order
parameter signaling the spontaneous breakdown of the discrete symmetry in this
metastable phase,
\beq\label{vevssb}
<\hat O> \;=\; \frac{2\,d\,\mu_1\,\phi_m}{L\,t^+_{\mu_1}(z_1)}\,t_1
\eeq

\section{\label{corfun} About correlation functions}

Having an exact solution, we can try to analytically compute the correlation functions
of the boundary theory.

The derivatives of the potential (\ref{vt}), (\ref{ABC}) are
\bea\label{Dvt}
V'(t)&=& \left\{\begin{array}{l} A_1\,t \\A_2\,(t-t_M)\\A_3\,(t-1) \end{array}\right.
\qquad;\qquad
V''(t)= \left\{\begin{array}{l}A_1>0 \\A_2<0\\A_3>0\end{array}\right.\\
V^{(n+3)}(t) &=&
(A_2 - A_1)\,\delta^{(n)}(t-t_1) +  (A_3 - A_2)\,\delta^{(n)}(t-t_2)\quad,\quad n=0,1,\ldots\nonumber
\eea
which will be used in the expansion
\beq
V'(t_b + \xi) = \sum_{n=0}^{\infty}\frac{1}{n!}\;V^{(n+1)}(t_b)\; \xi^n
\eeq
To compute correlation functions we refer the reader to the formalism sketched in Appendix B.
The equation of motion for a general field $t(x,z)$ is (\ref{eomgral})
and the solution is searched perturbatively via (\ref{Et1})-(\ref{eqm}).
According to (\ref{bcpert2}) the $m$-th approximation behaves as
\beq
\label{xiasymptotic}
z\to 0\quad:\quad \xi^{(m)}(x,z)\to\xi_{(0)}(x)\;z^{\Delta_-}\;\delta_{m,1}+
\xi_{(1)}^{(m)}(x)\;z^{\Delta_+}+\ldots
\eeq
with $\xi_{(1)}^{(m)}$ a functional of $\xi_{(0)}$.
Now, from equations (\ref{vev}), (\ref{Et1}), (\ref{xiexp}),
\beq\label{vevO}
<\hat O(x)>_{\phi_{(0)}} = \frac{\Delta_+ - \Delta_-}{L}\;\phi_{(1)}(x) =
\frac{2\,\nu_1\,\phi_m}{L}\;\left(t_b{}_{(1)} + \sum_{m=1}^\infty \xi^{(m)}_{(1)}(x)\right)
\eeq
where we introduced from the region $z\to 0$
\beq
\label{nu}
\nu_1=d\,\mu_1=\frac{\Delta_+-\Delta_-}{2}
\eeq
From here the result used so far (see (\ref{vev1}), (\ref{vevssb})) follows
\beq
<\hat O(x)> = \frac{2\,\nu_1\,\phi_m}{L}\;t_b{}_{(1)}
\eeq
For $n>1$ we have from (\ref{vevO}) and (\ref{key}),
\beq
\label{npointcf}
<\hat O(x_1)\dots \hat O(x_n)>= \frac{2\,\nu_1}{L\,\phi_m{}^{n-2}}\;
\frac{\delta^{n-1}\,\xi^{(n-1)}_{(1)}(x_1)}{\delta\xi_{(0)}(x_2)\dots\delta\xi_{(0)}(x_n)}
\eeq
In the following we will specialize to the $T=0$ case.

\subsection{The 2-point correlator}

Defining the momentum space first order perturbations through
\beq
\xi^{(1)}(x,z)\equiv\int\frac{d^d k}{(2\pi)^d}\;e^{i k\cdot x}\;\tilde\xi^{(1)}(k,z)
\eeq
equation (\ref{eq1}) for $\tilde\xi^{(1)}$  becomes ($\hat k\equiv L\,\sqrt{k^2}$, $f'\equiv\partial f/\partial z$)
\beq
z^2\,\tilde\xi^{(1)}{}''(k,z) - (d-1)\,z\,\tilde\xi^{(1)}{}'(k,z) -
\left( \hat k^2\,z^2+ V''(t_b)\right)\,\tilde\xi^{(1)}(k,z)=0
\eeq
with $V''(t_b)= A_i $ for $i=1,2,3$
depending on the region. The solution is
\beq
\label{xi1ssb}
\tilde\xi^{(1)}(k,z)=z^{\frac{d}{2}}\left\{
\begin{array}{lcr}
A(k)\;\;K_{\nu_1}(\hat k z) + B(k)\;I_{\nu_1}(\hat k z)
& , &  0\leq z< z_1\\
C(k)\;I_{i\bar\nu_2}(\hat k z) + C^*(k)\;I_{-i\bar\nu_2}(\hat k z)& , & z_1<z<z_2\\
D(k)\;K_{\nu_3}(\hat k z) & , &  z_2 <z < \infty
\end{array}
\right.
\eeq
where $A(k)$ is normalized by (\ref{xiasymptotic}) to be
\beq
\label{Ak}
A(k) = \frac{\hat k^{\nu}}{2^{\nu -1}\,\Gamma(\nu)}\tilde\xi_{(0)}(k)
\eeq
The coefficients $B(k), C(k)$ and $D(k)$ are determined by imposing continuity of
$\tilde\xi^{(1)}(k,z)$ and its derivative at $z=z_1, z_2$. All we need for our purpose is
\beq
\label{ba}
\frac{B(k)}{A(k)} = -\;\frac{Im\left[w(K_{\nu_1},I_{i\bar\nu_2};\hat kz_1)w(K_{\nu_3}, I_{-i\bar\nu_2};\hat k z_2)\right]}
{Im\left[w(I_{\nu_1},I_{i\bar\nu_2};\hat kz_1)w(K_{\nu_3}, I_{-i\bar\nu_2};\hat k z_2)\right]}
\eeq
where
\beq
\label{wronskian}
w(f,g;x)\equiv f(x)\,g'(x) - g(x)\,f'(x)
\eeq
denotes the Wronskian of $f$ and $g$ at $x$.
The normalization (\ref{Ak}) determines the coefficient of $z^{\Delta_1^-}$ in the
expansion of (\ref{xi1ssb}) for $z\rightarrow 0$ to be $\tilde\xi_{(0)}(k)$, while the
coefficient of $z^{\Delta_1^+}$ is
\beq\label{xi11ssb}
\tilde\xi^{(1)}_{(1)}(k) = -\frac{\Gamma(1-\nu_1)}{\Gamma(1+\nu_1)}\,
\frac{\hat k^{2\nu_1}}{2^{2\nu_1}}\;\left( 1 - \frac{2}{\Gamma(\nu_1)\,\Gamma(1-\nu_1)}\,
\frac{B(k)}{A(k)}\right)\;\tilde\xi_{(0)}(k)
\eeq
Finally, the two-point function can be calculated from
(\ref{npointcf}):
\beq
<\hat O(x_1) \hat O(x_2)>= \frac{2\,\nu_1}{L}\;
\frac{\delta\,\xi^{(1)}_{(1)}(x_1)}{\delta\xi_{(0)}(x_2)}
\eeq
Defining the Fourier transform
\beq
\langle\hat O(x_1)\hat O(x_2)\rangle=\int\frac{d^dk}{(2\pi)^d}
e^{ik(x_1-x_2)}G_2(k)
\eeq
we get in our SSB solution case from (\ref{xi11ssb})
\beq
\label{g2ssb}
G_2^{SSB}(k) = -\frac{2\,\nu_1}{L}\;\frac{\Gamma(1-\nu_1)}{\Gamma(1+\nu_1)}\,
\frac{\hat k^{2\nu_1}}{2^{2\nu_1}}\;\left( 1 - \frac{2}{\Gamma(\nu_1)\,\Gamma(1-\nu_1)}\,
\frac{B(k)}{A(k)}\right)
\eeq
The second term in the parenthesis signals the deviation of the CFT. 
In fact in the conformal vacuum ($t_b=0$) the propagator reduces to the well known form
\beq\label{G2free}
G_2(k)= -\frac{2\,\nu_1}{L}\;\frac{\Gamma(1-\nu_1)}{\Gamma(1+\nu_1)}\;
\frac{\hat k^{2\nu_1}}{2^{2\nu_1}}\quad\longleftrightarrow\quad
\langle\hat O(x_1)\hat O(x_2)\rangle= \frac{2\,\nu_1}{L}\;
\frac{\Gamma(\Delta_+^{(1)})}{\pi^\frac{d}{2}\,\Gamma(\nu_1)}\;
\frac{1}{|x_1-x_2|^{2\,\Delta_+^{(1)}}}
\eeq
Let us now analyze the propagator (\ref{g2ssb}) in some limits.

We first expand (\ref{g2ssb}) for small $k$ (IR limit).
To calculate the numerator of (\ref{ba}), the leading terms of the Wronskians (\ref{wronskian})
are enough, but for the denominator these leading terms give zero due to the constraint
(\ref{z12cont}), so that the next terms are needed. For $\nu_3>1$ (this is automatically
true for $d>2$) the propagator has a pole,
\beq
k^2\to 0\quad:\quad G_2^{SSB}(k)\to \frac{8z_1^{-2\nu_1}\nu_1(1+\bar\nu_2^2)}{(\nu_1^2+\bar\nu_2^2)\left(\frac{1+\nu_3}{1-\nu_3}z_2^2-
\frac{1-\nu_1}{1+\nu_1}z_1^2\right)}\frac{1}{\hat k^2}
\eeq
and obviously a corresponding massless mode associated with it. 
This is the Goldstone mode associated to the SSB of the conformal invariance due to the vev 
deformation \cite{Skenderis:2002wp}.

Finally, at large $k$, i.e. in the UV, the SSB propagator approaches exponentially fast the
conformal one. The boundary QFT is thus conformal in the UV. What is relevant in this limit is just the parameter
that determines the potential close to $\phi=0$, i.e. $\mu_1$ (or equivalently $A_1$) only.

\begin{figure}[htb]
\begin{center}
\includegraphics[width=7.cm]{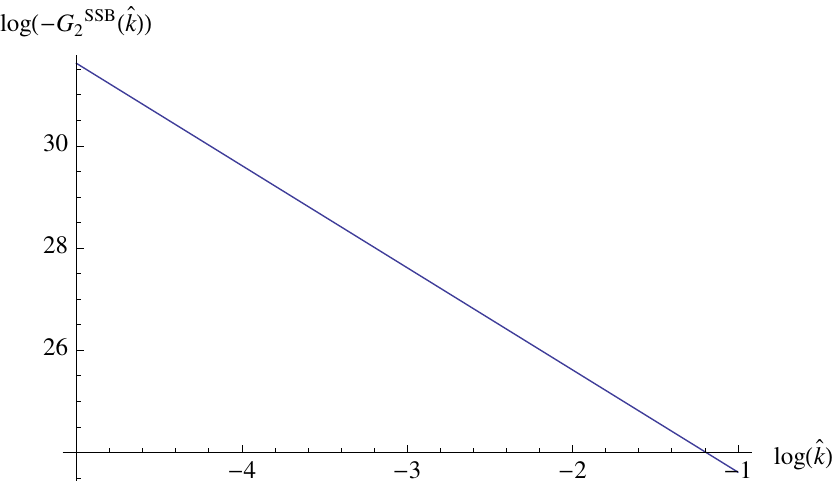}
\hspace{1cm}
\includegraphics[width=7.cm]{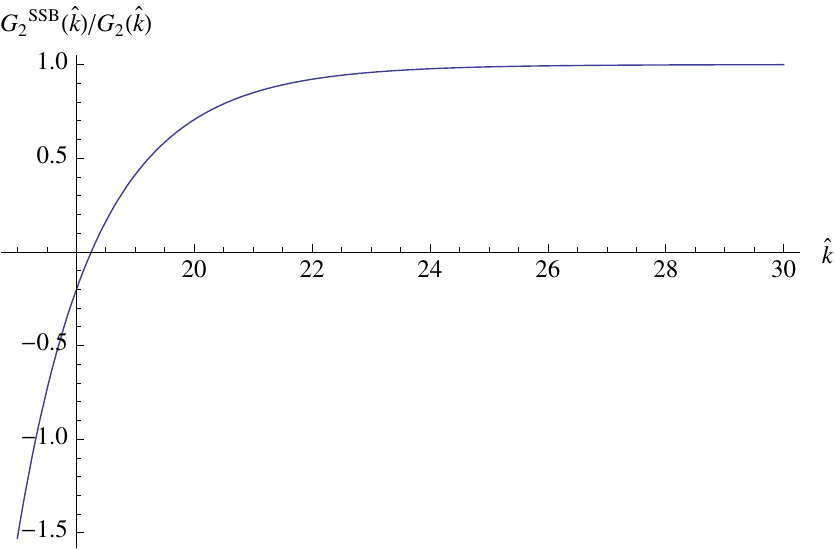}
\caption{\label{fig3} The behavior of the propagator in the SSB case for $T=0$ at small $\hat k$ (left) and large $\hat k$ (right) for
$\mu_1=1.1$, $\bar\mu_2=10.1$, $\mu_3=5.1$, $t_M=0.715$ ($k=0$). Notice the zero of the propagator for $\hat k\approx 18.25$.}
\end{center}
\end{figure}

The behavior of the propagator in the IR and UV is plotted on fig. \ref{fig3}.

Notice that the choice of $z_1$, which is not determined by the equation of motion, is arbitrary,
but the different choices generate different boundary QFT (different correlators). The meaning of
this $z_1$ is connected to the boundary in momentum space between the CFT in the UV region
($k>>1/z_1$) and the QFT of a massless mode in the IR region ($k<<1/z_1$). So it seems we have
a continuous family of theories that interpolate between these two limits. The massless mode is
a consequence of this flat direction: no energy is needed for the wall to move, so the excitation
in this direction is cost-free, i.e. with vanishing eigenvalue.

\subsection{The 3-point correlator}

For the 3-point correlation function we need to compute $\tilde\xi^{(2)}(x,z)$ from (\ref{Em}).
The solution for $m=2$ in (\ref{m>1}) reduces to
\beq
\xi^{(2)}(x,z)=\int d^dx'dz'\sqrt{det\,g_{ab}}\;G(x,z;x',z')\;\frac{1}{2}\;V^{(3)}(t_b(z'))\;
\left(\xi^{(1)}(x',z')\right)^2
\eeq
where $G(x,z;x',z')$ is the bulk-bulk propagator. Fortunately we do not need to compute it, since the
relevant part is just the $z^{\Delta_+}$ coefficient of $\xi^{(2)}(x,z)$ in the small $z$ expansion,
\beq
z\to 0\quad:\quad G(x,z;x',z')\to \frac{1}{2\,\nu_1\,L}\;z^{\Delta_1^+}\;K(x',z';x)
\eeq
where we introduced the bulk-boundary propagator
\beq
\label{bulkboundary}
K(x,z;x') =\int\frac{d^dk}{(2\pi)^d}\;e^{i k\cdot (x-x')}\;K(k,z)\quad,\quad
K(k,z)=\frac{\tilde\xi^{(1)}(k,z)}{\tilde\xi_{(0)}(k)}
\eeq
Using (\ref{m=1}) and (\ref{npointcf}) we get
\beq
\langle\hat O(x_1)\;\hat O(x_2)\;\hat O(x_3)\rangle=
\frac{1}{L^2\,\phi_m}\; \int d^dx\,\int dz\,\sqrt{\det g_{ab}}\;
V^{(3)}(t_b(z))\; \prod_{i=1}^3\,K(x,z;x_i)
\eeq
This same result could have been of course derived from Witten's diagrams recipe.

The case of conformal vacuum would give here a vanishing 3-point function, as
required by the unbroken $Z_2$ symmetry.

In our case we have from (\ref{Dvt})
\bea
V^{(3)}(t_b(z))&=&(A_2-A_1)\;\delta(t_b(z)-t_b(z_1))+(A_3-A_2)\;\delta(t_b(z)-t_b(z_2))\cr
&=&\frac{(A_2-A_1)}{\left|t_b'(z_1)\right|}\;\delta(z-z_1)+
\frac{(A_3-A_2)}{\left|t_b'(z_2)\right|}\;\delta(z-z_2)
\eea
The momentum space three-point correlation function from
\bea
\langle\hat O(x_1)\;\hat O(x_2)\;\hat O(x_3)\rangle&=&\int\frac{d^dk_1}{(2\pi)^d}
\int\frac{d^dk_2}{(2\pi)^d}\int\frac{d^dk_3}{(2\pi)^d}\;
(2\pi)^d\;\delta^d(k_1+k_2+k_3)\nonumber\\
&\times&e^{ik_1x_1}\,e^{ik_2x_2}\,e^{ik_3x_3}\;G_3(k_1,k_2,k_3)
\eea
remains thus without any integration:
\bea
L\;\phi_m\;G_3(k_1,k_2,k_3)&=&
\frac{(A_2-A_1)}{z_1^{d+1}\left|t_b'(z_1)\right|}\;K(k_1,z_1)\;K(k_2,z_1)\;K(k_3,z_1)\cr
&+&\frac{(A_3-A_2)}{z_2^{d+1}\left|t_b'(z_2)\right|}\;K(k_1,z_2)\;K(k_2,z_2)\;K(k_3,z_2)
\eea

The 3-point correlator has the right poles for $k_i^2=0$, as it should due to the
presence of the massless asymptotic state.

\section{Discussion and conclusions}

In this paper we have assumed that the tree level approximation is good enough. In fact here, contrary to
what happens in some string theory constructions, we do not have any large $N$ argument for this to be true.
What we did was just to solve perturbatively the classical bulk equation of motion with sources, which, by definition,
gives the tree order generating functional for connected diagrams. The next step would be to consider loops,
which is far beyond the scope of this paper. Notice however that this situation is similar to most models of
superconductivity that have appeared in the literature in recent years.

Another approximation we used is $\kappa\to 0$, i.e. no back-reaction. Once back-reaction (finite $\kappa$)
is introduced, no exactly solvable solutions are known, and only numerical analysis can be applied. Of course
in such a case considering a piece-wise quadratic potential is no more needed, since analytic results are anyway
unavailable. On the other side, although we will not attempt to do it here, there is hope to attack some issues in
the back-reacted system even without demanding numerical work. Namely, one could try to calculate the action
along the lines of subsection \ref{freeenergy} and appendix \ref{derivationeq} and check whether the free energy
gets lifted or not. Similar studies can be found in the literature, for example \cite{Skenderis:1999mm}, where BPS
type domain walls solutions in gauged supergravity context were considered. 
We remark that the non trivial solutions analyzed in this paper are not domain walls that 
describe renormalization group flows driven by deformation of the CFT by a relevant operator; they 
describe the CFT just in a non conformal vacuum.

We found out that the $Z_2$ symmetry breaking and preserving vacua are both equally possible at $T=0$. The two phases
can be co-existent, since they have the same free energy. This behavior changes at non-zero (high) temperature.
Although there exists a non-trivial solution also in this case, the free energy associated with this SSB vacuum is higher
than the trivial (symmetry preserving) one. This system thus seems to prefer symmetry restoration at high temperature.
It is however interesting that in many systems the non trivial phase disappears with the transition (for example in \cite{Hartnoll:2008vx},
\cite{Lugo:2010qq}), but not here neither in \cite{Aprile:2011uq} where examples from holographic models
of superconductors derived from gauged supergravity are given. It is thus reasonable to interpret this non-trivial solution
at $T>0$ found here as describing a metastable phase of symmetry non-restoration at high temperature
\cite{Weinberg:1974hy,Mohapatra:1979qt}, see for example \cite{Senjanovic:1998xc} for a review on this subject.

For the case $T=0$ we noticed that dilatation symmetry leads to a curious behavior: in order to get a non-trivial
solution to the equations of motion, a relation among the parameters of the potential is needed to be satisfied.
Although this means a strange fine-tuning for the bulk parameters, it has a physical effect for the boundary
theory, i.e. a massless mode. The existence of this mode is a consequence of dilatation symmetry: changing
the position of the domain wall between the two vacua costs no energy, and the quantum mechanical excitation
connected is obviously massless. In other words, dilatation symmetry is global, and the position of the wall breaks
it spontaneously. The massless mode is thus the Goldstone originating from it, i.e. the dilaton. 
Then one would expect the propagator at $T\ne 0$ not to have a pole at $k^2=0$, since there is no symmetry to start with. 
Although we were unable to solve the equations for the perturbations analytically in this case, the
absence of a pole seems reasonable: there is no constraint (\ref{tmk}) at $T\ne 0$, so
no reason to expect a magic cancelation of the leading term in $k^2\to 0$ of the denominator in (\ref{ba}).

The analysis of the previous section showed that in the UV ($k\to\infty$) limit the calculated correlation functions
approach their correspondent correlation functions of the conformal vacuum. What gets corrected is
the opposite, IR limit. And the theory here is a strongly interacting theory of a massless mode.

In this paper we limit ourselves to the 1, 2 and 3-point correlation functions. This is mainly due to simplicity. Higher
point correlation functions can be studied in a similar fashion, but involve the use of the bulk-bulk propagator, which,
although straightforward, is a bit more involved.
Also, two particular situations were avoided in this paper. One is the case of integer $\nu_1$, which gives
logarithmic corrections that complicate the analysis. The other case is $\nu_1$ in the conformal window
\cite{Breitenlohner:1982jf,Klebanov:1999tb}, for which
both $\Delta_\pm$ are positive. We plan to cover these and related topics in a follow-up publication.

\section*{Acknowledgments}
We would like to thank Jorge Russo, Mart\'in Schvellinger and Guillermo Silva for discussions, and
Giuseppe Policastro for correspondence.
This work has been supported in part by the Slovenian Research
Agency, and by the Argentinian-Slovenian programme BI-AR/09-11-006//
MINCYT-MHEST SLO/08/06. BB would like to acknowledge the Department of Physics of La Plata University, and ARL and MBS would like to acknowledge the J. Stefan Institute, Ljubljana, and ICTP, Trieste, for hospitality.

\appendix

\section{\label{derivationeq}Derivation of equation $(\ref{actionfinal})$}

Since we believe that (\ref{actionfinal}) is not completely obvious, we present a short derivation of it.

We want to simplify the expression for the action
\beq
\label{actioni}
I[t]\equiv\int_0^{z_h} \frac{dz}{z^{d+1}}\left(\frac{z^2}{2}(1-(z/z_h)^d)t'^2(z)+V(t(z))\right)
\eeq
for the solutions of the e.o.m.
For them we can replace one $z\,t'(z)$ above from the equation of motion
\beq
z\,t'=\frac{1}{d-1}\;\left((1-(z/z_h)^d)\;z^2\;t''-V'(t)-(z/z_h)^d\;z\;t'\right)
\eeq
Then the first term in (\ref{actioni}) can be rewritten as
\bea\label{first}
&&\int_0^{z_h} \frac{dz}{z^{d+1}}\;\frac{z^2}{2}(1-(z/z_h)^d)\;t'^2\cr
&=&\int_0^{z_h}\frac{dz}{z^{d+1}}\;\frac{(1-(z/z_h)^d)}{2\,(d-1)}\;z\;t'\;
\left((1-(z/z_h)^d)\;z^2\;t''-V'(t)-(z/z_h)^d\;z\;t'\right)\cr
&=&\int_0^{z_h}dz\;\frac{(1-(z/z_h)^d)^2}{2\,(d-1)\,z^{d-2}}\;
\left(\frac{t'^2}{2}\right)'-
\int_0^{z_h}dz\;\frac{(1-(z/z_h)^d)}{2\,(d-1)\,z^d}\;\frac{dV}{dz}\cr
&-&\int_0^{z_h}dz\;\frac{(z/z_h)^d\,(1-(z/z_h)^d)}{2\,(d-1)\,z^{d-1}}\;t'^2\cr
&=&\left(\frac{(1-(z/z_h)^d)^2}{2\,(d-1)\,z^{d-2}}\;\frac{t'^2}{2}\right)_0^{z_h}-
\int_0^{z_h}dz\;\frac{d}{dz}\left(\frac{(1-(z/z_h)^d)^2}{2\,(d-1)\,z^{d-2}}\right)\;
\frac{t'^2}{2}\cr
&-&\left(\frac{1-(z/z_h)^d}{2\,(d-1)\,z^d}\;V(t)\right)_0^{z_h}+
\int_0^{z_h}dz\;\frac{d}{dz}\left(\frac{(1-(z/z_h)^d)}{2\,(d-1)\,z^d}\right)V(t)\cr
&-&\int_0^{z_h}dz\;\frac{(z/z_h)^d\,(1-(z/z_h)^d)}{2\,(d-1)\,z^{d-1}}\;t'^2
\eea
where we used $t'dV/dt=dV/dz$ and integration by parts.
By rearranging and simplifying (\ref{first}) we get the following identity for any solution of the equation of motion $t(z)$:
\bea
&&d\,\int_0^{z_h}\;\frac{dz}{2\,z^{d-1}}\;(1-(z/z_h)^d)^2\;t'^2\cr
&=&\left[\frac{1-(z/z_h)^d}{z^d}\;\left(\frac{z^2}{2}\;(1-(z/z_h)^d)\;t'^2-V(t)\right)
\right]_{z=0}^{z=z_h}- d\,\int_0^{z_h}\;\frac{dz}{z^{d+1}}\;V(t)
\eea
Since
\beq
(1-(z/z_h)^d)^2=(1-(z/z_h)^d)-(z/z_h)^d\;(1-(z/z_h)^d)
\eeq
we get finally
\bea
I[t_b]&=&\left[\frac{(1-(z/z_h)^d)}{d\,z^d}\left(\frac{z^2}{2}\;(1-(z/z_h)^d)\;t_b'^2
-V(t_b)\right)\right]_{z=0}^{z=z_h}\cr
&+&\frac{1}{2}\int_0^{z_h}dz\;(z/z_h)^d\;\frac{1-(z/z_h)^d}{z^{d-1}}\;t_b'^2
\eea
for any solution to the equation of motion $t_b(z)$, which is nothing else than (\ref{actionfinal}).

\section{Formalism to compute correlation functions}
\bigskip

According to (\ref{genfunct}), to compute arbitrary $m$-points correlation functions we must obtain a solution to the equation of motion
\bea\label{eomgral}
0 &=& D^2(\phi) - U'(\phi)  \equiv \frac{\phi_m}{L^2}\; E[t]\cr
E[t] &=& (1-(z/z_h)^d)\;z^2\;t''(x,z) -(d-1+(z/z_h)^d)\;z\; t'(x,z)\cr
&+& L^2\;z^2\;\left(\frac{1}{1-(z/z_h)^d}\,
\partial^2_\tau t (x,z) +\nabla^2 t(x,z)\right) - V'(t)
\eea
with arbitrary boundary value $t(x,0)$ but finite $t(x,z_h)$. To this end we take the following route:

\begin{itemize}

\item We consider
\beq\label{Et1}
t(x,z) = t_b(z) + \xi(x,z)
\eeq
with $t_b(z)$ the background field solution of (\ref{eqfortgeneral}) dual to the boundary QFT.
The following expansion holds ($T=0$),
\bea
E[t_b + \xi] &=& \hat L [t_b]\xi - \sum_{n=2}^\infty\,\frac{1}{n!}V^{(n+1)}(t_b)\,\xi^n\cr
\hat L [t_b]\xi &\equiv& \left(z^2\,\partial_z^2 -(d-1)\,z\,\partial_z
+ L^2\,z^2\,\Box_x - V''(t_b(z))\right)\,\xi(x,z)
\eea

\item We introduce the expansion,
\beq\label{xiexp}
\xi(x,z) \equiv \sum_{m=1}^\infty\,\lambda^m\, \xi^{(m)}(x,z)
\eeq

where the parameter $\lambda$ is just to count powers and will be set to one at the end.

\item
We merge this expansion in the functional $E[t_b+\xi]$; we get,
\beq\label{expaneq}
E[t_b +\xi] = E[t_b + \sum_{m=1}^\infty\,\lambda^m\, \xi^{(m)}] \equiv \sum_{m=1}^\infty
\lambda^m\;E^{(m)}[\xi^{(1)},\dots,\xi^{(m)};t_b]
\eeq
where for the first three terms,
\bea\label{Em}
E^{(1)}[\xi^{(1)};t_b] &=&\hat L [t_b]\xi^{(1)}\cr
E^{(2)}[\xi^{(1)}, \xi^{(2)};t_b]&=&\hat L [t_b]\xi^{(2)} - \frac{1}{2}V^{(3)}(t_b)\,\left(\xi^{(1)}\right)^2\cr
E^{(3)}[\xi^{(1)},\xi^{(2)},\xi^{(3)};t_b]&=& \hat L [t_b]\xi^{(3)}
-V^{(3)}(t_b)\,\xi^{(1)}\,\xi^{(2)}- \frac{1}{6}V^{(4)}(t_b)\,\left(\xi^{(1)}\right)^3
\eea

\item

We ask for (\ref{Et1}) to be a solution to the equation of motion,
i.e. $r.h.s. (\ref{expaneq})=0$, by imposing $E^{(m)}=0$ for all $m$.
That is, we first solve,
\beq\label{eq1}
\hat L [t_b]\xi^{(1)} = 0
\eeq
and successively,
\beq\label{eqm}
\hat L [t_b]\xi^{(m)} = J^{(m)}[x,z;\xi^{(1)},\dots,\xi^{(m-1)};t_b]\quad,\quad m=2,3,\dots
\eeq
where the sources $J^{(m)}$ are read from (\ref{expaneq})
\footnote{
If $V(t)$ were strictly quadratic, $V^{(n+1)}(t)= 0\;,\; \forall\, n\geq 2$,
and then from (\ref{Em}) it follows $J^{(m)}=0\;,\; \forall\, m\geq 2$.
Then (\ref{eqm}) together with the boundary conditions  (\ref{bcpert2}) imply
$\xi^{(m)}(x,z) = 0\;,\; \forall\, m\geq 2$.
}, for $m=1,2,3$ explicitly from (\ref{Em}).

\end{itemize}

To solve the system we must impose the following boundary conditions:
for $z\rightarrow 0$ on the background solution they are,
\beq\label{bcback}
t_b(z) \rightarrow t_v + t_b{}_{(1)}\,z^{\Delta_+} + \dots
\eeq
where $t_v$ is a critical point of the potential, while that for the perturbation,
\beq\label{bcpert1}
\xi(x,z) \rightarrow \xi_{(0)}(x)\,z^{\Delta_-} +\dots +
\xi_{(1)}(x)\,z^{\Delta_+} + \dots
\eeq
where $\xi_{(0)}(x)$ is a fixed, but arbitrary source.
This will be implemented in the field expansion for $z\to 0$ as
\beq\label{bcpert2}
z^{-\Delta_-}\,\xi^{(1)}(x,z) \rightarrow \xi_{(0)}(x)\qquad;\qquad z^{-\Delta_-}\,\xi^{(m)}(x,z)
\rightarrow 0\quad,\quad m=2,3,\dots
\eeq
Furthermore, at the horizon $z\to z_h$ we should ask that,
\beq
z^{-\Delta_-}\,\xi^{(m)}(x,z)<\infty\qquad ,\qquad m=1,2,\dots
\eeq
When $T=0$ the horizon is at $z_h=\infty$ and this condition enforces $\xi$ to go to zero fast enough.
When $T>0$ the horizon is at $z=z_h=1$ and so this is just a finiteness condition for the $\xi^{(m)}$'s, i.e. for $\xi$.

Equation (\ref{eq1}) is solved by,
\beq\label{m=1}
\xi^{(1)}(x,z) =  \int_{M_d}\; d^d x'\;K(x,z;x')\;\xi_{(0)}(x')
\eeq
where the bulk-boundary propagator matrix $K$ obeys,
\beq
\hat L[t_b]\, K(x,z;x') = 0\qquad;\qquad K(x,z;x')\longrightarrow
z^{\Delta_-}\;\delta^d(x,x')\quad,\quad z\rightarrow 0
\eeq
For $m>1$ instead, equations (\ref{eqm}) are not homogeneous and we can write the solution in the way,
\beq\label{m>1}
\xi^{(m)}(x,z) =  \int\; d^d x'\,dz'\,\sqrt{\det g_{ab}}\;G(x,z;x',z')\; J^{(m)}(x',z')
\eeq
where the bulk-bulk propagator  $G$ obeys,
\beq
\hat L[t_b]\, G(x,z;x',z') = \delta_g^{d+1}(x,z;x',z')
\qquad;\qquad z^{-\Delta_-}\;G(x,z;x',z')\longrightarrow 0\quad,\quad z\rightarrow 0
\eeq
Now, from (\ref{m=1}) it is obvious that $\xi^{(1)}$ is linear in $\xi_{(0)}$.
Also, direct inspection of (\ref{m>1}) shows that the following key property holds,
\beq\label{key}
\xi^{(m)} \propto \left(\xi_{(0)}\right)^m\quad,\quad m=1,2,\dots
\eeq
With all this baggage we can return to our task of computing the $m$-point correlation functions of $\phi(x,z) = \phi_m\,t(x,z)$.
From (\ref{xiasymptotic}), (\ref{vevO}) and (\ref{key}) we get,
\beq
<\hat O(x_1)\dots\hat O(x_m)> =\frac{2\,\nu_1}{L}
\frac{\delta^m\varphi_{(1)}^{(m-1)}(x_1)}{\delta\varphi_{(0)}(x_2)\ldots
\delta\varphi_{(0)}(x_m)}
\eeq

This procedure generates formally the well-known ``Witten diagrams", and can be
straightforwardly extended to any QFT defined by its gravity dual.

\end{document}